\documentclass[aps,prl,preprint,groupedaddress]{revtex4}
\usepackage{graphicx}
\usepackage{amsmath,amsfonts,amssymb}
\usepackage{color}
\begin{document}

\title{Restoring Time Dependence into Quantum Cosmology}

{\sf{Honorable Mentioned essay - Gravity Research Foundation 2012}}
\bigskip

\author{Aharon Davidson}
\author{Ben Yellin}

\affiliation{Physics Department, Ben-Gurion University of the Negev,
Beer-Sheva 84105, Israel \bigskip}


\begin{abstract}
	Mini superspace cosmology treats the scale factor $a(t)$, the
	lapse function $n(t)$, and an optional dilation field $\phi(t)$ as
	canonical variables.
	While pre-fixing $n(t)$ means losing the Hamiltonian constraint,
	pre-fixing $a(t)$ is serendipitously harmless at this level.
	This suggests an alternative to the Hartle-Hawking approach,
	where the pre-fixed $a(t)$ and its derivatives are treated as
	explicit functions of time, leaving $n(t)$ and a now mandatory
	$\phi(t)$ to serve as canonical variables.
	The naive gauge pre-fix $a(t)=const$ is clearly forbidden, causing
	evolution to freeze altogether, so pre-fixing the scale factor, say
	$a(t)=t$, necessarily introduces explicit time dependence into the
	Lagrangian.
	Invoking Dirac's prescription for dealing with constraints,
	we construct the corresponding mini superspace time dependent
	total Hamiltonian, and calculate the Dirac brackets, characterized
	by $\{n,\phi\}_D\neq 0$, which are promoted to commutation
	relations in the quantum theory.
	\\  \\{\sf{Email: davidson@bgu.ac.il, yellinb@bgu.ac.il}}
\end{abstract}

\maketitle
Let our starting point be the simple general relativistic (GR) action
\begin{equation}
	{\cal S}=-\frac{1}{16\pi G}\int\left({\cal R}
	+2\Lambda\right)\sqrt{-g}~d^4 x ~,
	\label{GR}
\end{equation}
involving a positive cosmological constant $\Lambda$ or, alternatively,
its minimal dilaton gravity (DG) variant
\begin{equation}
	{\cal S}=-\frac{1}{16\pi}
	\int\left(\phi {\cal R}+V(\phi)\right)\sqrt{-g}~d^4 x ~.
	\label{DG}
\end{equation}
Adding a Brans-Dicke kinetic term to the latter is optional, but even
in its absence, for the sake of simplicity, the dilation field $\phi(x)$
is fully dynamical, subject to the Klein-Gordon equation
\begin{equation}
	g^{\mu\nu}\phi_{;\mu\nu}=
	\frac{1}{3}\left(\phi V^{\prime}(\phi)-2V(\phi)\right)
	\equiv V_{eff}^{\prime}(\phi)~.
\end{equation}

Our interest lies with cosmology, so let the corresponding line
element take the form
\begin{equation}
	ds^2=-n^2(t)dt^2+a^2(t)
	\left(\frac{dr^2}{1-kr^2}+r^2 d\Omega^2\right)~,
	\label{FLRW}
\end{equation}
with $a(t),n(t)$ denoting the scale factor and the lapse function,
respectively, and let $k=-1,0,1$ classify the maximally symmetric
3-subspaces.
Notably, the time re-definition symmetry $t\rightarrow f(t)$
is still there, with the standard gauge choice $n(t)=1$, for example,
defining the FLRW cosmic time.
Recalling the fact that $n(t)$ plays a major role in canonical
quantum gravity \cite{ADM}, one has to be careful regarding the stage
at which this gauge freedom can be harmlessly exercised.

Given the cosmological line element eq.(\ref{FLRW}), one can
integrate the spatial dimensions out of the above action to
arrive at the reduced mini superspace \cite{HH} action
$\displaystyle{\tilde{{\cal S}}=\int {\cal L}~dt}$.
After subtracting a total derivative, to get rid of the accompanying
$\ddot{a}$-term, the mini superspace Lagrangian $\cal L$ takes the
explicit form
\begin{equation}
	{\cal L}=-\left( \frac{1}{6}a^3 V(\phi)-k a \phi \right) n
	-\left( \dot{a} \phi+a \dot{\phi}\right)\frac{a \dot{a}}{n}~.
	\label{L}
\end{equation} 
Note that one can always return to the simpler GR case by setting
$\phi=G^{-1}$, and choosing the constant potential
$V(\phi)=2\Lambda G^{-1}$.
At this stage our discussion trifurcates:

\smallskip
$\bullet$ Had we pre-fixed the lapse function,
substituting (say) $n(t)=1$ into eq.(\ref{L}) before conducting the
variation, we could not have recovered, starting from
${\cal L}(a,\dot{a},\phi,\dot{\phi})$, the correct classical solution.
Instead, already for the GR action, one would encounter a superfluous
matter density contribution, as is evident from the (integrated)
Friedmann equation
\begin{equation}
	\frac{\dot{a}^2+k}{a^2}-\frac{\Lambda}{3}=\frac{E}{a^3}~.
\end{equation}
This is a reflection the fact that the Hamiltonian constraint has been
simply thrown away.

\smallskip
$\bullet$ Appreciating the cruicial role played by the lapse function
at the mini superspace model, Hartle and Hawking \cite{HH} have suggested,
in the spirit of canonical quantum gravity, to add $n(t)$ to the list of
canonical variables, so that ${\cal L}(n,\dot{n},a,\dot{a},\phi,\dot{\phi})$.
The fact that the momentum
$\displaystyle{p_n=\frac{\partial {\cal L}}{\partial \dot{n}}}$ happens
to vanish constitutes a primary constraint.
For the latter to stay a constant of motion, consistency requires
its Poisson's brackets with the Hamiltonian to weekly vanish, that is
\begin{equation}
	\frac{dp_n}{dt}=\left\{p_n,{\cal H}\right\}_P+
	\frac{\partial p_n}{\partial t} \approx 0.
\end{equation}
To be more specific, demonstrating again for the simpler GR action
eq.(\ref{GR}), the Hamiltonian ${\cal H}=p_n \dot{n}+p_a \dot{a}-{\cal L}$
turns out to be proportional to $n$, thereby leading to the famous
Hamiltonian constraint
\begin{equation}
	\left\{p_n,{\cal H}\right\}_P=-\frac{{\cal H}}{n}
	=\frac{1}{a}\left(\frac{p_a^2}{4}
	+k a^2-\frac{\Lambda a^4}{3}\right) =0~.
\end{equation}
At the classical level, this ensures a vanishing 'mechanical energy' $E=0$.
At the semi quantum mechanical level \cite{HH,L,V}, applying the standard
operator assignment $\displaystyle{P_a \rightarrow -i\frac{\partial}{\partial a}}$,
and up to the usual order ambiguity, one encounters the Wheeler-DeWitt
equation \cite{WdW}
\begin{equation}
	{\cal H}(p_a,a)\psi(a)=0 ~.
\end{equation}
Frustratingly, the cosmological wave function $\psi(a)$, which is supposed
to govern the quantum mechanical evolution of the universe, happens to
be time independent.

\smallskip
\noindent $\bullet$ Once $n(t)$ is elevated to the level of a legitimate
canonical variable, the question is whether the canonical role of $a(t)$
can be relaxed?
In other words, can one harmlessly pre-fix $a(t)$ before conducting
the variation?
Serendipitously, the answer is in the affirmative.
It is crucial to notice, however, that the naive choice $a(t)=const$ is forbidden
(a tenable choice, for example, is $a(t)=t$), as evolution gets frozen altogether, so
that \emph{pre fixing the scale factor necessarily introduces explicit time
dependence into the Lagrangian}, i.e.
${\cal L}(n,\dot{n},\phi,\dot{\phi},t)$.
In turn, being ready to deviate from the Hartle-Hawking approach,
this opens the door for restoring time dependence into the cosmological
wave function.

For the GR action eq.(\ref{GR}), with $a(t)$ pre-fixed, the Euler Lagrange
equation, which is reduced now to
\begin{equation}
	\frac{\partial {\cal L}}{\partial n}=
	-\frac{1}{3}\Lambda a^3+ka+\frac{a\dot{a}^2}{n^2}=0~,
	\label{reduced}
\end{equation}
does give rise to the correct algebraic equation for $n(t)$ (rather than
to a differential equation for $a(t)$).
The Hamiltonian constraint, while consistently reproducing
eq.(\ref{reduced}), is momentum free, and therefore falls short of
supporting a differential wave equation.
This unfortunate situation is going to be changed once the dilation
field enters the game.
In this respect, the optional $\phi(t)$ becomes mandatory,
and hence it is the DG-action eq.(\ref{DG}) which gets our attention.

Given the Lagrangian eq.(\ref{L}), with pre-fixed $a(t)$ and its derivatives
treated as explicit functions of time, the corresponding momenta
$\displaystyle{p_n=\frac{\partial {\cal L}}{\partial \dot{n}}}$ and
$\displaystyle{p_{\phi}=\frac{\partial {\cal L}}{\partial \dot{\phi}}}$
fail to determine the velocities $\dot{n}$ and $\dot{\phi}$.
This, in turn, gives rise to the two primary constraints
\begin{equation}
	\O_1 = p_n \approx 0, \quad
	\O_2=p_{\phi}+\frac{a^2\dot{a}}{n} \approx 0 ~.
	\label{O12}
\end{equation}
It is crucial to notice that the Poisson brackets of these two constraints
does not vanish.
To be specific, we have
\begin{equation}
	\{\O_1,\O_2\}_P=\frac{a^2 \dot{a}}{n^2} \neq 0 ~.
	\label{Poisson}
\end{equation}

The time dependent and strikingly momentum free naive Hamiltonian
${\cal H}=p_n \dot{n} +p_{\phi} \dot{\phi}- {\cal L}$ is given by
\begin{equation}
	{\cal H}(n,\phi,t)= \left(\frac{1}{6}a^3 V(\phi)-k a\phi \right)n+\frac{a\dot{a}^2 \phi}{n}~.
\end{equation}
However, as argued by Dirac \cite{Dirac}, the Hamiltonian defined in this way
is not uniquely determined, and one may add to it any linear
combination of the $\O$'s, which are zero, and go over to
\begin{equation}
	{\cal H}^{\star}={\cal H}+\sum_i u_i \O_i ~.
	\label{star}
\end{equation}
Consistency then requires the constraints be constants of motion,
and as such, they must weakly obey
\begin{equation}
	\frac{d\O_i}{dt}=\{\O_i,{\cal H}\}_P+\sum_j u_j \{\O_i,\O_j \}_P
	+\frac{\partial\O_i}{\partial t}\approx 0 ~,
	\label{constraints}
\end{equation}
which generalizes the previously discussed Hamiltonian constraint.
Applying the latter to our two primary constraints $\O_{1,2}$, one
of which depends explicitly on time, we use the non-vanishing Poisson
brackets eq.(\ref{Poisson}) to calculate the coefficients $u_i$.
We find
\begin{eqnarray}
	&& u_1 (n,\phi,t)=\frac{\displaystyle{k-\frac{1}{6}a^2 V^{\prime}(\phi)}}{a\dot{a}}n^3
	+\left(\frac{\dot{a}}{a}+\frac{\ddot{a}}{\dot{a}}\right)n ~, \\
	&& u_2 (n,\phi,t)=\left(\frac{1}{6}V(\phi)-\frac{k\phi}{a^2}
	\right)\frac{a n^2}{\dot{a}} -\frac{\dot{a}\phi}{a} ~.
\end{eqnarray}
Using Dirac's terminology \cite{Dirac}, substituting the $u_{1,2}(n,\phi,t)$
coefficients into eq.(\ref{star}) then constitutes the so-called total
Hamiltonian ${\cal H}_T (p_n,n,p_{\phi},\phi,t)$.
As a re-check, one can easily verify that 
\begin{equation}
	\frac{\partial {\cal H}_T}{\partial p_n}
	=u_1=\frac{dn}{dt}~, ~~
	\frac{\partial {\cal H}_T}{\partial p_{\phi}}
	=u_2=\frac{d\phi}{dt}
\end{equation}
match the correct Lagrangian equations of motion stemming
from ${\cal L}(n,\dot{n},\phi,\dot{\phi},t)$, and furthermore
note that with $u_{1,2}$ substituted, eqs.(\ref{constraints}) are
automatically satisfied.
In other words, ${\cal H}_T (p_n,n,p_{\phi},\phi,t)$ is all we need,
without any additional constraints attached.

Given the non-vanishing Poisson brackets eq.(\ref{Poisson}),
telling us that our two primary constraints are in fact second-class, the
corresponding Dirac brackets take the form
\begin{equation}
	\{A,B\}_D=\{A,B\}_P+
	\frac{n^2}{a^2 {\dot a}}\epsilon_{ij}\{A,\O_i\}_P\{\O_j,B\}_P ~.
	\label{Db}
\end{equation}
It has been argued \cite{deLeon} that the Dirac brackets formula
is supposed to get modified in the presence of time depended
constraints.
To stay on the safe side, owing to the fact that only one constraint
is now time dependent, see eq.({\ref{O12}), we have verified that
eq.(\ref{Db}) acquires no such further modification (if at all) in the
present case.
If we wish to quantize the ${\cal H}_T (p_n,n,p_{\phi},\phi,t)$ theory
we should compute the Dirac brackets between all of our momenta
and coordinates so that these may be promoted to commutation
relations.
Doing so, we find
\begin{equation}
	\begin{array}{clc}
  	\{n,p_n\}_D=0~, & 
	\quad\displaystyle{\{n,\phi\}_D=-\frac{n^2}{a^2{\dot a}}}~,&
	 \quad\{n,p_{\phi}\}_D=0~, \\
 	\{p_n,\phi\}_D=0~,&
	 \quad\{p_n,p_{\phi}\}_D=0 ~,&
	 \quad\{\phi,p_{\phi}\}_D=1 ~.	
	\end{array}
\end{equation}
Of particular interest are the somewhat surprising facts that:
\begin{itemize}
	\item The two canonical coordinates $n$ and $\phi$ no longer
	commute,
	suggesting perhaps an underlying non-commutative geometry.
	\item $n$ and $p_n$ do commute, as if they are classical objects, but
	such an unusual feature could have actually been expected recalling
	the special role played by the $p_n$=0 constraint.
	\item And in particular, $\phi$ and $p_{\phi}$ continue to
	stay a canonical pair even under the Dirac bracket formalism.
\end{itemize}
Elevating the above Poisson bracket to the level of commutation
relations, and up to various order ambiguities, one is led to a
time dependent cosmological Schrodinger equation of the generic
type
\begin{equation}
	{\cal H}_T \psi (n,\phi,t)
	=-i\frac{\partial}{\partial t}\psi (n,\phi,t)~.
\end{equation}
A reduced variant of the latter can be constructed by substituting
the constraints, in the form of
$\displaystyle{n=-\frac{a^2 \dot{a}}{p_{\phi}}}$ accompanied by $p_n =0$
into the total Hamiltonian ${\cal H}_T (p_n,n,p_{\phi},\phi,t)$, giving
rise to the so-called reduced Hamiltonian ${\cal H}_R(p_{\phi},\phi,t)$.
This leads, without any lose of generality, to a much simpler
Schrodinger equation, namely
\begin{equation}
	{\cal H}_R \psi (\phi,t)
	=-i\frac{\partial}{\partial t}\psi (\phi,t)~,
\end{equation}
which takies full advantage of the fact that
$\displaystyle{\phi,p_{\phi}=i\frac{\partial}{\partial\phi}}$
form a canonical pair.
The metric can then always be reconstructed via
$\displaystyle{n=-\frac{a^2 \dot{a}}{p_{\phi}}}$.

Finally, arriving at the quantum stage, note the special pre-fix
$a(t)=a_C (t)$, with $a_C (t)$ being the classical scale factor solution.
\emph{This way, the FLRW cosmic time returns to play a role in
quantum cosmology.}
A neat pedagogical example for $a_C (t)$ in DG gravity is provided
by the quadratic potential
\begin{equation}
	V(\phi)=\lambda\left(\phi-\frac{1}{G} \right)^2
	+2G \Lambda \phi^2
	\quad \Longrightarrow \quad
	V_{eff}(\phi)=\frac{\lambda}{3G}
	\left(\phi-\frac{1}{G} \right)^2 ~.
\end{equation}
A few remarks regarding this potential are in order:
(i) The VEV $\displaystyle{\langle\phi\rangle=\frac{1}{G}}$, which
is stable for $\lambda>0$, is accompanied by a constant Ricci
curvature solution ${\cal R}=-4\Lambda$, corresponding to a
cosmological constant $\Lambda$.
(ii) Counter intuitively, $V_{eff}(\phi)$ is $\Lambda$-independent.
The evolution of the dilaton is controlled by $\Lambda$ only indirectly,
via the space-time metric, and
(iii) In the cosmological case, especially during creation, stability is
clearly not an issue.
On the contrary, a linear potential for which $\lambda=-2G\Lambda$,
implying $\lambda<0$ for $\Lambda>0$, namely
\begin{equation}
	V(\phi)=4\Lambda \left(\phi -\frac{1}{2G}\right)~,
\end{equation}
will certainly do, and can
be used as the simplest example.
In fact, associated with the linear potential, is the classical solution
\begin{equation}
	n^2 (t)=\frac{{\dot a}^2}
	{\displaystyle{\frac{\Lambda}{3}a^2-k+\frac{s}{a^2}}}~,
\end{equation}
with the $s$-term ($s$ being a constant of integration) interpreted
as radiation density.
$a_C (t)$ is then the solution of the differential equation associated
with $n(t)=1$.
All curvature scalars, including the Kretschmann scalar, are smooth
and non-singular at the instance of creation for which the denominator
vanishes.

\acknowledgments{
Special thanks to BGU president Prof. Rivka Carmi for the
kind support.}

\end{document}